\documentclass{cjaa}                   

\usepackage{graphicx}                   
\input{epsf.sty}                        
\input{psfig.sty}                       

\begin{document}

   \title{Classifying the zoo of ultraluminous X-ray sources
}

   \volnopage{Vol.0 (200x) No.0, 000--000}      
   \setcounter{page}{1}          

   \author{Roberto Soria
      \inst{1}\mailto{}
   \and Mark Cropper
      \inst{1}
   \and Christian Motch
      \inst{2}
      }
   \offprints{R. Soria}                   

   \institute{MSSL, University College London, 
	Holmbury St Mary, RH5 6NT, United Kingdom\\
             \email{Roberto.Soria@mssl.ucl.ac.uk}
        \and
             Observatoire de Strasbourg, 11, rue de l'Universit\'{e}, 
		F-67000 Strasbourg, France\\
          }

   \date{Received~~2004 July 15; Accepted 2004 September 6}

   \abstract{
Ultraluminous X-ray sources (ULXs) are likely to include  
different physical types of objects. 
We discuss some possible subclasses, reviewing 
the properties of a sample of ULXs recently observed 
by {\it Chandra} and {\it XMM-Newton}. 
Sources with an isotropic X-ray luminosity up to a few times 
$10^{39}$ erg s$^{-1}$ are consistent with ``normal'' 
stellar-mass X-ray binaries (mostly high-mass X-ray binaries 
in star-forming regions). Higher black hole (BH) masses ($\approx 50$--$100 M_{\odot}$) 
may be the end product of massive stellar evolution 
in peculiar environments: they may explain ULXs with 
luminosities $\approx 1$--$2 \times 10^{40}$ erg s$^{-1}$.
Only a handful of ULXs require a true intermediate-mass BH 
($M \ga 500 M_{\odot}$).
Finally, a small subclass of ULXs shows flaring or rapid variability 
in its power-law spectral component.
   \keywords{accretion, accretion disks --- black hole physics 
--- galaxies: individual: NGC\,4559 --- X-ray: galaxies --- X-ray: stars}
   }

   \authorrunning{R. Soria, M. Cropper, \& C. Motch}            
   \titlerunning{The ULX zoo}  

   \maketitle

%
%
\section{Introduction}           
\label{sect:intro}
ULXs are defined as
point-like, accreting sources with apparent 
isotropic luminosities $L_{\rm x} > 10^{39}$ erg s$^{-1}$, 
not including supermassive BHs in AGN and quasars.
Masses, ages and mechanisms of formation of the 
accreting objects are still unclear, as is the geometry 
of emission. In fact, ULXs are likely to include 
different physical classes of sources. 
At least four scenarios have been suggested:  
{\it a)} intermediate-mass BHs (Colbert \& Mushotzky 1999), 
with masses $\sim 10^2$--$10^3 M_{\odot}$; 
{\it b)} normal BH X-ray binaries ($M \la 20 M_{\odot}$) 
with mild geometrical beaming, during phases of super-Eddington 
mass accretion (King et al.~2001); {\it c)} 
microquasars with a relativistic jet pointing towards us 
(Fabrika \& Mescheryakov 2001; K\"{o}rding et al.~2002); 
{\it d)} BH X-ray binaries with super-Eddington emission 
from inhomogeneous disks (Begelman 2002).
These different scenarios also predict different 
duty cycles for the X-ray-bright phases, 
thus leading to different predictions for the location and 
total number of such systems (active or quiescent) in a galaxy.

ULXs have been detected in many nearby spiral, irregular 
and elliptical galaxies; however, most sources brighter 
than $\approx 2 \times 10^{39}$ erg s$^{-1}$ 
are located in star-forming galaxies, associated 
with young populations (Irwin et al.~2004). 
In the few cases when stellar counterparts have been identified 
(Soria et al.~2004a and references therein),
they have masses $\approx 15$--$30 M_{\odot}$.  
X-ray ionized emission nebulae 
have been found around many ULXs (Pakull \& Mirioni 2002; 
Kaaret et al.~2004): 
in some cases, they provide evidence against 
beaming (Holmberg II X-1: Pakull \& Mirioni 2002); 
in other cases, they suggest anisotropic 
emission (IC 342 X-1: Roberts et al.~2003).
The presence of a young, massive star cluster near 
a ULX (as suggested for the ULXs in the Antennae: 
Zezas et al.~2002) is consistent with the formation 
of an intermediate-mass BH from merger processes 
in a cluster core.
However, for most of the sources, multi-band 
observations so far have not been able to rule out  
any of the alternative scenarios.



\section{A case study: different types of ULXs in NGC\,4559}
\label{sect:Obs}
Located at a distance of $\approx 10$ Mpc, 
the late-type spiral NGC\,4559 (Type SAB(rs)cd) 
has been observed by {\it XMM-Newton} and {\it Chandra} 
on various occasions between 2001 and 2003 (Cropper et al.~2004; 
Soria et al.~2004a; Roberts et al.~2004). It hosts four sources 
detected at $L_{\rm x}\ga 10^{39}$ erg s$^{-1}$ 
on at least one occasion. Here we briefly summarize the properties 
of these four sources: we argue that they may represent four different 
classes of ULXs. 

\subsection{Intermediate-mass BHs in nuclear star clusters?}

A variable ($L_{\rm x} \approx 1.5 \times 10^{39}$ erg s$^{-1}$ 
in 2002 March; $L_{\rm x} < 10^{38}$ erg s$^{-1}$ in 2001 June), 
point-like X-ray source coincides with the nucleus 
of NGC\,4559. Bulgeless Scd galaxies do not contain a supermassive BH 
in their nucleus; this is in agreement with the observed relations between 
the nuclear BH mass and the luminosity and velocity dispersion of the bulge 
(Magorrian 1998; Ferrarese \& Merritt 2000). 
Instead, many of them contain a bright, 
massive nuclear star cluster ($M \sim 10^6 
M_{\odot}$, $L \sim 10^6$--$10^7 L_{\odot}$; B\"{o}ker et al.~2004), 
with a complex history of intermittent star formation. 
The nuclear star cluster in NGC\,4559 has $M_B \approx -12$, 
$M_I \approx -13$. It is marginally resolved in the {\it HST}/PC image, 
with a full-width half maximum of $\approx 5$ pc.
The Local-Group galaxy M\,33 provides another example 
of a ULX ($L_{\rm x} \approx 1.5 \times 10^{39}$ 
in the $0.5$--$10$ keV band: Dubus \& Rutledge 2002)
in the nuclear star cluster ($M_B = -10.2$: Kormendy \& McClure 1993) 
of a bulgeless late-type spiral, without a supermassive BH 
($M_{\rm BH} < 1500 M_{\odot}$: Gebhardt et al.~2001).

Theoretical and observational studies have investigated 
the possible formation of intermediate-mass BHs 
in the core of old globular clusters (Miller \& Hamilton 2002) 
or of young super star clusters (Portegies Zwart et al.~2004; 
G\"{u}rkan et al.~2004).
We suggest that nuclear star clusters in late-type spirals 
may offer another natural environment for these objects, 
and can be used as a test for BH formation and galaxy merger models.


\subsection{Bright high-mass X-ray binaries?}

A ULX with $L_{\rm x} \approx 2$--$3 \times 10^{39}$ erg s$^{-1}$ 
is located in the inner disk of NGC\,4559, along a spiral arm, in a region 
of young star formation. Its spectrum is well fitted 
by a disk-blackbody model with color temperature 
$kT_{\rm in} \approx 0.88$--$1.07$ keV (in 2003 and 2001 respectively), 
and inner disk radius $\approx 110/(\cos \theta)^{1/2}$ km (in 2003) 
or $\approx 85/(\cos \theta)^{1/2}$ km (in 2001), 
where $\theta$ is the viewing angle of the disk. 
Assuming an efficiency $\sim 0.1$, 
the mass accretion rate is $\sim$ a few $\times 10^{19}$ 
g s$^{-1}$, i.e., $\sim$ a few $\times 10^{-7} M_{\odot}$ yr$^{-1}$.
Color temperature and luminosity are consistent with 
the emission from a standard disk around a 
$\approx 15$--$25 M_{\odot}$ BH, accreting close to its Eddington 
limit, in a high/soft state.
Hence, this source can be considered a normal stellar-mass 
BH X-ray binary, probably with an OB donor star filling 
its Roche lobe. 
The BH mass is similar to what is inferred 
for the most massive BH candidates in the Milky Way: for example, 
$M = (14.0\pm 4.4) M_{\odot}$ for the BH in GRS 1915$+$105 
(Harlaftis \& Greiner 2004), 
and $M \approx 16$--$32 M_{\odot}$ for the BH in Cyg X-1 
(Ziolkowski 2004; Gies \& Bolton 1986).
It is likely that the majority of ULXs 
with $10^{39} \la L_{\rm x} \la 10^{40}$ 
erg s$^{-1}$ fall in this category.

\subsection{Young, massive BHs from stellar evolution?}

The brightest ULX in NGC\,4559, X7, 
has a $0.3$--$10$ keV luminosity $\approx 2$--$3 \times 10^{40}$ erg s$^{-1}$, 
and an extrapolated bolometric luminosity $> 5 \times 10^{40}$ erg s$^{-1}$ 
(Cropper et al.~2004). Hence, it is too bright 
to be a ``normal'' stellar-mass remnant, 
if we assume strictly Eddington-limited isotropic emission.
We have discussed the case for three possible mass ranges 
(Soria et al.~2004a). A stellar-mass BH 
($M \la 20 M_{\odot}$) would require either strong, collimated 
beaming (it would effectively be a microblazar) 
or strongly super-Eddington emission. 
A $\sim 50$--$100 M_{\odot}$ BH accreting 
from a $15$--$25 M_{\odot}$ supergiant companion would require 
mildly anisotropic emission (geometrical beaming 
of a factor $\sim 5$--$10$; King 2004) or mildly super-Eddington 
luminosity. An intermediate-mass BH ($M \ga 500 M_{\odot}$) 
would be consistent with isotropic, sub-Eddington emission.

Our X-ray spectral and timing study (Cropper et al.~2004) 
has shown the presence of a very soft thermal component 
($kT_{\rm in} \approx 0.12$--$0.15$ keV), 
and a break in the power-density-spectrum at 0.03 Hz. 
These findings are not consistent with a microblazar 
scenario, and suggest a BH mass $\ga 50 M_{\odot}$.
The optical data 
indicate an association with a young region of massive star formation 
at the edge of the galactic disk (age $< 30$ Myr). 
They also rule out an association between the ULX and any star clusters; 
hence, they rule out at least one mechanism of formation for 
an intermediate-mass BH. Other formation processes 
have been proposed (e.g., from Pop-III 
stars), but they are more difficult to reconcile with a young age 
and disk location. Therefore, the X-ray and optical data 
together are more consistent with a young $50$--$100 M_{\odot}$ BH 
originating from massive stellar evolution 
and accreting from an OB supergiant companion (Soria et al.~2004a).

Such a massive remnant would require a progenitor star 
at least twice as massive. One can only speculate that 
such massive stars, not observed in the Milky Way, 
may exist, and evolve into massive remnants, 
in other galactic environments. NGC\,4559 X7 is located 
in a region of active star formation triggered by a satellite 
galaxy collision (Soria et al.~2004a), and in a metal-poor 
environment. These two conditions are also a characteristic 
of many other bright ULXs. We speculate 
that such environments can affect the equation 
of state of the gas, and the balance of heating and cooling, 
reducing the fragmentation process in the collapse 
of a molecular cloud core, and leading to the formation 
of more massive stars (Spaans \& Silk 2000). 
Low metal abundance 
has an effect at later stages of the stellar 
evolution, by reducing the mass-loss rate 
in the radiatively-driven wind (Pakull \& Mirioni 2002).
This leads to a more massive stellar core, 
which may then collapse into a more massive BH.

\subsection{NGC\,4559 X10: still a mystery}

Another source with isotropic luminosity 
$L_{\rm x} \ga 10^{40}$ erg s$^{-1}$ (NGC\,4559 X10)
is located in the inner galactic disk, in an inter-arm region, 
away from young OB associations or H{\footnotesize {II}} regions.
Its X-ray spectrum is consistent with 
a simple power-law (photon index $\Gamma \approx 2$), 
without a significant disk component. 
Its nature is still a mystery, and may well be different from
the three classes of ULXs described above. 
Possible speculations include an isolated intermediate-mass BH 
accreting from a molecular cloud (R. Mushotzky, priv. comm.; see also 
Krolik 2004), or a stellar-mass microblazar.
Further {\it HST} observations of this source are scheduled for 
2005 March; we are also planning to carry out radio observations.

\section{True intermediate-mass BHs in colliding systems?}

Colliding and tidally-interacting systems seem to offer 
the most favorable environment for ULXs. 
For example, ULXs have been found in the Antennae (Zezas et al.~2002), 
the Cartwheel ring (Gao et al.~2003), 
M\,82 (Griffiths et al.~2000), the Mice (Read 2003), 
the M\,81 group dwarfs (Wang 2002). 
NGC\,4559 X7 is also associated with a minor collisional 
event: we have suggested (Soria et al.~2004) 
that the initial perturbation responsible for the expanding 
wave of star formation was caused by a satellite dwarf galaxy 
crossing the gas-rich outer disk of NGC\,4559.

Another example of a ULX in a colliding system 
was found at the intersection of the collisional ring 
of NGC\,7714 (mostly old stars, with negligible gas or star formation) 
with the gas-rich bridge between NGC\,7715 and NGC\,7714 (Soria \& Motch 2004). 
The bridge contains a string of young super star clusters, though none 
is found near the ULX position. Hydrodynamical simulations 
of the interaction between the two galaxies (Struck \& Smith 2003) 
suggest that the connecting bridge consists of multiple components, 
and that the most recent star-formation episode (responsible 
for the string of young clusters) was triggered by their interaction, 
which shocked or compressed the gas. It was also suggested 
that part of the gas in the bridge is currently infalling 
onto NGC\,7714, impacting the outer disk at approximately 
the ULX location.
The ULX luminosity reached $\approx 6 \times 10^{40}$ 
erg s$^{-1}$ in 2002 December ($0.3$--$12$ keV band), 
implying an estimated bolometric luminosity 
$\ga 1.5 \times 10^{41}$ erg s$^{-1}$. If isotropic, it requires 
an accreting BH with a mass $\ga 500 M_{\odot}$, 
ruling out its origin from single stellar evolution processes.
The X-ray spectrum is featureless, and in the absence of an accurate 
{\it Chandra} position and of an identified optical counterpart, 
none of the proposed scenarios can be ruled out at present.

At least three other bright ULXs fall in this category:
a source in NGC\,2276 (tidally interacting 
with NGC\,2300), with a luminosity $\approx 1.1 \times 10^{41}$ 
erg s$^{-1}$ in the $0.5$--$10$ keV band (Davis \& Mushotzky 2004); 
a ULX in the colliding Cartwheel galaxy, with $L_{\rm x} 
\approx 1 \times 10^{41}$ erg s$^{-1}$ (Gao et al.~2003); 
and a variable source in M\,82 with a $L_{\rm x} 
\approx 1$--$9 \times 10^{40}$ erg s$^{-1}$ (Matsumoto et al.~2001). 
The last source has been interpreted as an intermediate-mass BH 
($M \sim 800$--$3000 M_{\odot}$), 
from numerical simulations 
of runaway collisions in cluster cores 
(Portegies Zwart et al.~2004). 
This explanation seems more difficult 
to reconcile with the ULXs in NGC\,2276 and NGC\,7714.


\section{ULXs with short-term variability}

X-ray observations may provide two methods to constrain 
the mass of the accreting BH. When a soft thermal component 
is present in the spectrum, fitting a disk model provides 
the color temperature at the inner disk boundary; 
hence, one can infer the radius of the last stable orbit 
and the BH mass.
For example, thermal components at $kT \approx 0.15$ keV 
have been found in many ULXs, and were initially used 
as evidence for an intermediate-mass BH (Miller et al.~2003). 
However, this method is based on highly questionable 
assumptions on the accretion flow structure, 
and the mass estimates thus derived are probably unreliable. 

Alternatively, X-ray timing observations can provide 
the characteristic variability time-scale in the X-ray emitting 
region; this is also assumed to be related to the inner-disk size 
and hence to the BH mass. A linear correlation appears to exist between 
the mass of an accreting BH and a characteristic break frequency 
in its power-density-spectrum (e.g., McHardy et al.~2004; 
Markowitz et al.~2003).
Evidence for characteristic breaks has been found 
in a few sources, most notably for the ULX in the metal-poor 
starburst dwarf galaxy NGC\,5408 ($L_{\rm x} \approx 10^{40}$ erg s$^{-1}$: 
Soria et al.~2004b; Kaaret et al.~2003).
The position of the break, at $\approx 2.5$ mHz, suggests 
a mass at least one order of magnitude larger than Cyg X-1. 
Moreover, the flux variability 
is associated with a spectral change. The X-ray spectrum 
can be interpreted as the sum of a constant or slowly-variable 
soft thermal component ($kT_{\rm bb} \approx 0.12$ keV) and 
a rapidly variable power-law component 
(photon index $\Gamma \approx 2.7$), 
responsible for the flaring behaviour (Soria et al.~2004b). 
The source gets harder at higher fluxes, at the flare peaks.

ULXs with short-term variability are comparatively rare 
(Roberts et al.~2004).
M\,74 X-1 has a $0.3$--$10$ keV luminosity 
varying from $\approx 5 \times 10^{38}$ to $\approx 1.2 \times 10^{40}$
erg s$^{-1}$, over timescales of a few $\times 10^3$ s, 
and is harder at higher fluxes (Krauss et al.~2004). 
The flaring behaviour has been attributed to the power-law component, 
while the thermal component is less variable. 
Similar behaviour is exhibited by NGC\,6946 X-11 ($\approx 2.5 \times 10^{39}$
erg s$^{-1}$: Roberts \& Colbert 2003), 
with a rapidly-varying power-law component (timescale $\la 350$ s).
There are at least three possible interpretations 
for the spectral variability properties of ULXs 
such as NGC\,5408 X-1, M\,74 X-1 and NGC\,6946 X-11.
The first scenario is that the power-law flaring may be due to 
intermittent ejections and variable emission at the base of a jet. 
If so, these ULXs would be microquasars or microblazars; the accreting 
BH would still have to be rather massive ($M \ga 40 M_{\odot}$ for 
NGC\,5408 X-1) to explain the non-beamed thermal disk emission, 
which provides a lower limit to the mass. 
Alternatively, the variability could be due to magnetic 
reconnections in the disk corona (Reeves et al.~2002).
Finally, in the framework of the two-component 
accretion model (Chakrabarti \& Titarchuk 1995), 
the power-law variability can be due to rapid 
changes in the sub-Keplerian (halo) accretion rate, 
$\dot{M}_{\rm h} \approx 0.5$--$1 \dot{M}_{\rm Edd}$, 
at a constant disk accretion rate $\dot{M}_{\rm d}$, 
when $\dot{M}_{\rm d} \approx 0.1$--$0.5 \dot{M}_{\rm Edd}$ 
(corresponding to the transition regime 
between the ``canonical'' hard and soft states).

\section{Conclusions}
\label{sect:conclusion}
ULXs are likely to contain 
different physical types of sources, in terms of BH mass, 
age, mechanism of formation and geometry of emission. 
We have briefly discussed some possible subclasses, reviewing 
the properties of a sample of ULXs recently observed by {\it Chandra} and 
{\it XMM-Newton}. Sources with an isotropic X-ray luminosity up to a few times 
$10^{39}$ erg s$^{-1}$ are consistent with bright X-ray binaries 
(in particular, young high-mass X-ray binaries in a star-forming region) 
containing a stellar-mass BH, in the same mass range as those 
inferred for Milky Way BH candidates. 
BH masses $\approx 50$--$100 M_{\odot}$ are required to explain 
a group of sources with X-ray luminosities $\approx 1$--$2 \times 10^{40}$ 
erg s$^{-1}$. We speculate that they, too, could be the end product 
of single stellar evolution, in metal-poor environments.
Only a handful of sources so far are strong candidates 
for the intermediate-mass BH class ($M \ga 500 M_{\odot}$).
Coalescence of smaller bodies in the cores 
of young super star clusters and nuclear star clusters  
are possible mechanisms of formation for intermediate-mass BHs; 
this scenario may apply to the brightest ULX in M\,82 
(Portegies Zwart et al.~2004).
However, most of the brightest ULXs do not reside 
in clusters, and may have been formed via other processes.
Finally, we have reviewed a sub-sample of ULXs which exhibit 
short-term variability and flaring in their Comptonised (power-law) 
emission component, and we have briefly discussed possible 
interpretations.


While we expect the number of ULXs in the high-mass X-ray binaries 
subclass to be proportional to the star-formation rate, 
there is mounting evidence that the brightest sources 
are preferentially found in colliding or tidally interacting 
systems. It is not yet clear whether star formation 
triggered by molecular cloud collisions 
can allow the formation of very massive stars, 
or whether there is another mechanism at play there: 
for example, the accretion or infall 
of primordial BHs from the galactic halo towards gas-rich parts 
of a galaxy, or the accretion and subsequent tidal stripping 
of satellite dwarfs with a nuclear intermediate-mass BH.
Multiwavelength observations will be necessary 
to disentangle the various subclasses of ULXs: 
for example, radio detections would identify microquasars 
and microblazars; infra-red surveys might allow us to find 
the reprocessed radiation from anisotropic X-ray sources not oriented 
along our line of sight (i.e., sources that would not 
be identified as ULXs); optical studies of X-ray ionized 
nebulae can constrain the total luminosity 
from the central sources.

\begin{acknowledgements}
We thank Miriam Krauss, Craig Markwardt, Richard Mushotzky, Manfred Pakull, 
Andrew Read, Jason Stevens, Doug Swartz and Kinwah Wu 
for useful discussions. 
\end{acknowledgements}

\label{lastpage}


\begin{thebibliography}{99}

\bibitem[2002]{beg}
Begelman, M. C. 2002, ApJ, 568, L97

\bibitem[2004]{bok}
B\"{o}ker, T., Sarzi, M., McLaughlin, D. E., et al.~2004, AJ, 127, 105 

\bibitem[1995]{cha}
Chakrabarti, S.,  \& Titarchuk, L. G. 1995, ApJ, 455, 623

\bibitem[1999]{col}
Colbert, E. J. M., \& Mushotzky, R. F. 1999, ApJ, 519, 89

\bibitem[2004]{crop}
Cropper, M. S., Soria, R., Mushotzky, R. F., et al.~2004, MNRAS, 349, 39

\bibitem[2004]{dav}
Davis, D. S., \& Mushotzky, R. F. 2004, ApJ, 604, 653

\bibitem[2002]{dub}
Dubus, G., \& Rutledge, R. E. 2002, MNRAS, 336, 901

\bibitem[2001]{fabrika}
Fabrika, S., \& Mescheryakov, A. 2001, in: High Angular Resolution in Astronomy, 
ed. R. Schilizzi et al. (ASP Publication) (astro-ph/0103070)

\bibitem[2000]{fer}
Ferrarese, L., \& Merritt, D. 2000, ApJ, 539, L9

\bibitem[2003]{gao}
Gao, Y., Wang, Q. D., Appleton, P. N., \& Lucas, R. A. 2003, 
	ApJ, 596, L171


\bibitem[2001]{geb}
Gebhardt, K., Lauer, T. R., Kormendy, J., et al. 2001, AJ, 122, 2469

\bibitem[1986]{gies}
Gies, D. R., \& Bolton, C. T. 1986, ApJ, 304, 371

\bibitem[2000]{grif}
Griffiths, R. E., Ptak, A., Feigelson, E. D., et al.~2000, Science, 290, 1325

\bibitem[2004]{gur}
G\"{u}rkan, M. A., Freitag, M., \& Rasio, F. A. 2004, ApJ, 604, 632

\bibitem[2004]{harl}
Harlaftis, E. T., \& Greiner, J. 2004, A\&A, 414, L13

\bibitem[2004]{irw}
Irwin, J. A., Bregman, J. N., \& Athey, A. E. 2004, ApJ, 601, L143

\bibitem[2003]{kaar1}
Kaaret, P., Corbel, S., Prestwich, A. H., Zezas, A.~2003, Science, 299, 365

\bibitem[2004]{kaar2}
Kaaret, P., Ward, M. J., \& Zezas, A. 2004, MNRAS, 351, L83

\bibitem[2004]{king1}
King, A. R. 2004, MNRAS, 347, L18

\bibitem[2001]{king2}
King, A. R., Davies, M. B., Ward, M. J., Fabbiano, G., 
\& Elvis, M. 2001, ApJ, 552, L109


\bibitem[2002]{kord}
K\"{o}rding, E., Falcke, H., \& Markoff, S. 2002, A\&A, 382, L13

\bibitem[1993]{kor}
Kormendy, J., \& McClure, R. D. 1993, AJ, 105, 1793



\bibitem[2004]{kra}
Krauss, M. I., Kilgard, R. E., Garcia, M. R., Roberts, T. P., 
\& Prestwich, A. H. 2004, ApJ, in press

\bibitem[2004]{kra}
Krolik, J. H. 2004, ApJ, in press (astro-ph/0407285)

\bibitem[1998]{mag}
Magorrian, J., et al. 1998, AJ, 115, 2285


\bibitem[2003]{mark}
Markowitz, A., Edelson, R., Vaughan, S., et al.~2003, ApJ, 593, 96

\bibitem[2001]{mat}
Matsumoto, H., Tsuru, T. G., Koyama, K., et al.~2001, ApJ, 547, L25

\bibitem[2004]{mch}
McHardy, I. M., Papadakis, I. E., Uttley, P., Page, M. J., 
\& Mason, K. O. 2004, MNRAS, 348, 783

\bibitem[2001]{mill1}
Miller, J. M., Fabbiano, G., Miller, M. C., \& Fabian, A. C. 2003, ApJ, 585, L37

\bibitem[2002]{mill2}
Miller, M. C., \& Hamilton, D. P. 2002, MNRAS, 330, 232

\bibitem[2002]{pak}
Pakull, M. W., \& Mirioni, L. 2002, to appear 
	in the proceedings of the symposium 'New Visions 
	of the X-ray Universe', 26-30 November 2001, 
	ESTEC, The Netherlands (astro-ph/0202488)

\bibitem[2004]{port1} 
Portegies Zwart, S. F., Baumgardt, H., Hut, P., et al.~2004, Nature, 428, 724

\bibitem[2003]{read}
Read, A. M. 2003, MNRAS, 342, 715

\bibitem[2002]{ree}
Reeves, J. N., Wynn, G., O'Brien, P. T., \& Pounds, K. A. 2002, MNRAS, 336, L56


\bibitem[2003]{rob1}
Roberts, T. P., \& Colbert, E. J. M. 2003, MNRAS, 341, L49


\bibitem[2003]{rob2}
Roberts, T. P., Goad, M. R., Ward, M. J., \& Warwick, R. S. 2003, MNRAS, 342, 709

\bibitem[2004]{rob3}
Roberts, T. P., Warwick, R. S., Ward, M. J., \& Goad, M. R. 2004, 
MNRAS, 349, 1193 (also, Erratum: 1994, MNRAS, 350, 1536)

\bibitem[2004]{sor1}
Soria, R., Cropper, M. C., Pakull, M., Mushotzky, R. F., \& Wu, K. 2004a, MNRAS, in press

\bibitem[2004]{sor2}
Soria, R., \& Motch, C. 2004, A\&A, 422, 915

\bibitem[2004]{sor3}
Soria, R., Motch, C., Read, A. M., \& Stevens, I. R. 2004b, A\&A, 423, 955 

\bibitem[2000]{spa}
Spaans, M., \& Silk, J. 2000, ApJ, 538, 115

\bibitem[2003]{str}
Struck, C., \& Smith, B. J. 2003, ApJ, 589, 157



\bibitem[2002]{wang}
Wang, Q. D. 2002, MNRAS, 332, 764

\bibitem[2002]{zez}
Zezas, A., Fabbiano, G., Rots, A. H., 
	\& Murray, S. S. 2002, ApJ, 577, 710

\bibitem[2004]{ziol}
Ziolkowski, J.~2004, ChJAA, submitted

\end{thebibliography}
\end{document}